\documentclass[acmsmall,screen]{acmart}
\usepackage{tabularx}  
\AtBeginDocument{%
  }

\setcopyright{acmlicensed}
\copyrightyear{2018}
\acmYear{2018}
\acmDOI{XXXXXXX.XXXXXXX}

\acmJournal{JACM}
\acmVolume{37}
\acmNumber{4}
\acmArticle{111}
\acmMonth{8}

\usepackage{graphicx}

\usepackage{amsmath}
\usepackage{amsthm}

\usepackage{ragged2e}
\usepackage{wasysym}
\usepackage{tabularx}  
\usepackage{multicol}
\usepackage{multirow}

\usepackage{subcaption} 

\usepackage{listings,framed,xcolor}
\definecolor{lgray}{rgb}{0.95,0.95,0.95}
\usepackage{float}
\usepackage{multirow}
\usepackage{mathrsfs}
\usepackage{algorithmic}
\usepackage[ruled,linesnumbered]{algorithm2e}
\SetKwRepeat{Do}{do}{while}

\usepackage{enumitem}

\usepackage{rotating}
\usepackage{balance}
\usepackage{threeparttable}
\usepackage{makecell}
\usepackage{pdflscape}

\usepackage{graphicx}

\usepackage[export]{adjustbox}

\usepackage{stfloats}
\usepackage{verbatim}
\usepackage{setspace}
\usepackage{threeparttable}
\usepackage{supertabular,booktabs}
\usepackage{rotating}
\usepackage{longtable}
\usepackage{supertabular}
\usepackage{pifont}
\usepackage{xcolor,pict2e}

\usepackage{xstring}
\usepackage{seqsplit}

\usepackage{lettrine}
\usepackage{hyperref}
\hypersetup{
    colorlinks=true,
    linkcolor=blue,    
    citecolor=blue,    
    urlcolor=black      
}
\makeatletter
\renewcommand\@cite[2]{%
    {\color{blue}[#1]}
}

\renewcommand\p@subfigure{\thefigure} 
\makeatother

\newcommand{\finding}[2]{%
\begin{tcolorbox}[
  breakable,
  enhanced,
  width=\linewidth,
  boxrule=0.7pt,          
  colframe=black, 
  colback=blue!5,         
  top=0.5pt,bottom=0.5pt,left=0.5pt,right=0.5pt,
  before skip=2.3pt, after skip=2.3pt,
  pad at break*=1mm,
]
\textbf{Answer to RQ#1:} #2
\end{tcolorbox}%
}

\makeatother
\usepackage[hyphenbreaks]{breakurl}
\usepackage{fancyhdr}

\allowdisplaybreaks[4] 

\usepackage{tcolorbox}
\tcbuselibrary{breakable} 
\usepackage{xcolor}
\usepackage{soul}   

\newcounter{ChenhuiCommentCounter}
\newcommand{\textttsplie}[1]{\texttt{\seqsplit{#1}}}

\definecolor{lightpink}{rgb}{1.0, 0.87, 0.87}
\definecolor{lightpurple}{rgb}{0.94, 0.85, 0.94}
\definecolor{lightgreen}{rgb}{0.88, 1.0, 0.88}
\definecolor{lightyellow}{rgb}{1.0, 1.0, 0.88}
\definecolor{lightblue}{rgb}{0.88, 0.94, 1.0}
\tcbuselibrary{skins,breakable,listings}

\newcounter{DaveCommentCounter}
\newcommand{\ddu}[1]{
    \stepcounter{DaveCommentCounter}
    \textcolor{blue}{\textit{/**Dave's comment [\arabic{DaveCommentCounter}]: I don't understand the intended meaning in the next sentence. Please revise/delete/explain. **/}}
}

\newcommand{\dns}[1]{
    \stepcounter{DaveCommentCounter}
    \textcolor{blue}{\textit{/**Dave's comment [\arabic{DaveCommentCounter}]: I'm not sure that I have captured the intended meaning in the next sentence. Please check/confirm. **/}}
}

\newcounter{AbelCommentCounter}
\newcounter{RubingCommentCounter}
\begin{document}
\title{LLM-Assisted Model-Based GUI Testing for Vue.js Web Applications}

\author{Tao Li}
\email{3220007015@student.must.edu.mo}
\orcid{0009-0001-7413-9692}
\affiliation{
  \institution{School of Computer Science and Engineering, Macau University of Science and Technology}
  \city{Macao SAR}
  \country{China}
  \postcode{999078}
}
\email{abelli@buaa.edu.cn}
\affiliation{
  \institution{School of Artificial Intelligence (Institute of Artificial Intelligence), Beihang University}
  \city{Beijing}
  \postcode{100191}
  \country{China}
}

\author{Chenhui Cui}
\email{3230002105@student.must.edu.mo}
\orcid{0009-0004-8746-316X}
\affiliation{
  \institution{School of Computer Science and Engineering, Macau University of Science and Technology}
  \city{Macao SAR}
  \country{China}
  \postcode{999078}
}

\author{Rubing Huang}
\email{rbhuang@must.edu.mo}
\orcid{0000-0002-1769-6126}
\affiliation{
  \institution{School of Computer Science and Engineering, Macau University of Science and Technology}
  \city{Macao SAR}
  \country{China}
  \postcode{999078}
}
\affiliation{
  \institution{Macau University of Science and Technology Zhuhai MUST Science and Technology Research Institute}
  \city{Zhuhai}
  \state{Guangdong Province}
  \country{China}
  \postcode{519099}
}

\author{Dave Towey}
\email{dave.towey@nottingham.edu.cn}
\orcid{0000-0003-0877-4353}
\affiliation{
  \institution{School of Computer Science, University of Nottingham Ningbo China}
  \city{Ningbo}
  \state{Zhejiang}
  \country{China}
}

\author{Shikai Guo}
\email{shikai.guo@dlmu.edu.cn}
\orcid{0000-0002-8554-6365}
\affiliation{
  \institution{School of Information Science and Technology, Dalian Maritime University}
  \city{Dalian}
  \postcode{116026}
  \country{China}   
}

\author{Lei Ma}
\email{ma.lei@acm.org}
\orcid{0000-0002-8621-2420}
\affiliation{
  \institution{School of Computer Science, University of Tokyo}
  \city{Tokyo}
  \country{Japan}
  \postcode{113-0033}
}
\affiliation{
  \institution{Department of Electrical and Computer Engineering, University of Alberta}
  \city{Edmonton}
  \country{Canada}
  \postcode{T6G 2R3}
}


\begin{abstract}
    Vue.js is a popular framework for building modern web applications. As Vue.js functionality and tooling support grow, ensuring its reliability (through automated testing) is becoming increasingly important. Although model-based testing has been successfully used to automate graphical user interface (GUI) testing on other platforms, its application to Vue.js remains challenging: Transition candidates, which are spread across router configurations and single-file components (SFCs), must be concretized and normalized into an executable page transition graph (PTG) for testing. To address this, we propose the \textsc{LLMVue} framework, which uses a large language model (LLM) to generate a PTG from Vue.js source code. \textsc{LLMVue} infers component hierarchies and route transitions, merging them into a unified PTG across multiple SFCs. We evaluated \textsc{LLMVue} on a collection of ten open-source Vue.js projects from GitHub, using GPT-4o as the LLM backbone. The constructed graphs demonstrate high precision and recall, with low graph edit distance. \textsc{LLMVue}-guided testing also significantly improves the coverage and exploration efficiency, compared to a random exploration baseline (with the same time constraints). To the best of our knowledge, this is the first use of LLMs for model-based GUI testing of Vue.js applications using source-level PTG extraction.
\end{abstract}



\received{18 June 2026}


\maketitle

\section{Introduction}
\label{sec:introduction}

As modern web technologies become more widely adopted, ensuring the reliability of large-scale web applications has become an important issue in software development~\cite{zhang2024towards,alian2024semantic,alian2025autoe2e,balsam2025web}.
Vue.js~\cite{vuejs} has become one of the most popular frameworks for modern web development~\cite{10mostpopularfrontendframeworks}, noted for its lightweight and reactive component architecture~\cite{nelson2018getting,yerburgh2018testing}.
In Vue.js applications, user interface (UI) pages are organized around client-side routing:
A router maintains a routing table that maps uniform resource locator (URL) paths to view components,  updating the rendered view when the active route changes.
A consequence is that the reliability of Vue.js applications is closely tied to the correctness of client-side navigation and page transitions. 
As the scale of Vue.js applications grows, automated testing has become an essential process for ensuring reliability.

\begin{figure*}[!t]
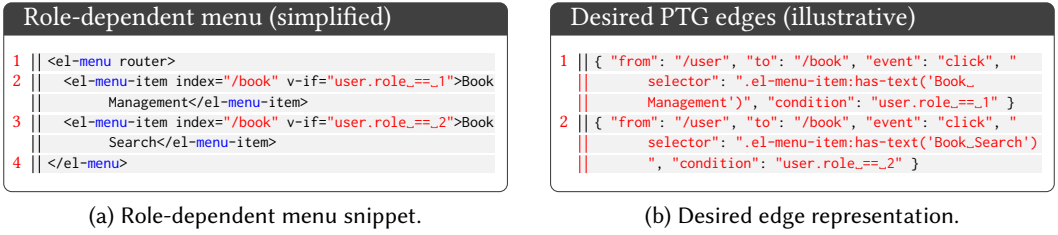

\vspace{-0.6em}
\centering

\begin{minipage}[t]{0.48\linewidth}
\vspace{0pt}
\begin{tcblisting}{
  enhanced,
  breakable,
  colback=white,
  title={Role-dependent menu (simplified)},
  boxrule=0.4pt,
  left=2.2mm,right=1.2mm,top=0.5mm,bottom=0.5mm,
  boxsep=0.4mm,
  before skip=0pt, after skip=0pt,
  listing only,
  listing options={
    basicstyle=\ttfamily\tiny\linespread{0.95}\selectfont,
    breaklines=true,
    breakatwhitespace=false,
    columns=fullflexible,
    keepspaces=true
  }
}
<el-menu router>
  <el-menu-item index="/book" v-if="user.role == 1">Book Management</el-menu-item>
  <el-menu-item index="/book" v-if="user.role == 2">Book Search</el-menu-item>
</el-menu>
\end{tcblisting}
\vspace{-0.2em}
\subcaption{Role-dependent menu snippet.}
\label{fig:mot-left}
\end{minipage}
\hfill
\begin{minipage}[t]{0.48\linewidth}
\vspace{0pt}
\begin{tcblisting}{
  enhanced,
  breakable,
  colback=white,
  title={Desired PTG edges (illustrative)},
  boxrule=0.4pt,
  left=2.2mm,right=1.2mm,top=0.5mm,bottom=0.5mm,
  boxsep=0.4mm,
  before skip=0pt, after skip=0pt,
  listing only,
  listing options={
    basicstyle=\ttfamily\tiny\linespread{0.95}\selectfont,
    breaklines=true,
    breakatwhitespace=false,
    columns=fullflexible,
    keepspaces=true
  }
}
{ "from": "/user", "to": "/book", "event": "click", "selector": ".el-menu-item:has-text('Book Management')", "condition": "user.role == 1" }
{ "from": "/user", "to": "/book", "event": "click", "selector": ".el-menu-item:has-text('Book Search')", "condition": "user.role == 2" }
\end{tcblisting}
\vspace{-0.2em}
\subcaption{Desired edge representation.}
\label{fig:mot-right}
\end{minipage}

\vspace{-0.6em}
\caption{Role-dependent navigation in Vue.js.}
\label{fig:motivating-example}
\vspace{-0.8em}
\end{figure*}

Model-based testing (MBT) \cite{memon2007event,yuan2009generating,chen2024model,dincturk2014model,van2015crawl} has proven its effectiveness for graphical user interface (GUI) testing.
It uses an explicit navigation model, typically a page transition graph (PTG), to capture transitions among UI pages and to guide action selection and test-case generation.
Despite substantial progress in PTG construction and PTG-based GUI testing \cite{chen2024model}, systematic transition modeling for Vue.js applications remains limited~\cite{vueSfcSpec}.
Vue.js introduces framework-specific features that complicate cross-component PTG construction:
(1) The navigation logic is spread across router configurations and single-file components (SFCs)\footnote{The official definition of a Vue single-file component (SFC), a ``\texttt{.vue}'' file, is available here: \url{https://vue-loader.vuejs.org/spec.html}.}. 
It can also be intertwined with layout components and nested routing.
(2) Many navigation actions are encapsulated within UI-library components (e.g., Element UI/Plus), which can hide the underlying clickable elements and routing APIs.
(3) Menus and navigation actions are often protected against certain actions by mutually exclusive runtime conditions (e.g., authentication state or permission checks), causing different transitions to be exposed only under specific states.
Unlike traditional multi-page web applications, Vue.js SPAs often encode navigation intent across router definitions, SFC templates/scripts, shared layouts, and UI-library abstractions. 
As a result, recovering an executable cross-component navigation model is not a direct extension of conventional web crawling or URL-based modeling.

To support model-based GUI testing on Vue.js, we construct an executable cross-component PTG from Vue.js source code by extracting transition candidates and using them to drive PTG-guided testing.
However, this approach introduces two challenges:

\textbf{(1) Candidate extraction for PTG construction, from cross-component Vue.js source code, prioritizing high recall.}
The navigation intent is distributed across router configurations, SFC templates/scripts, and shared layout components, and is often wrapped by UI library components. 
We therefore need a conservative static extraction that emphasizes high recall and collects plausible route-to-route transition candidates from these distributed sources.

\textbf{(2) Refining and normalizing static candidates into an executable PTG to guide testing.}
Static candidates are often incomplete or ambiguous (e.g., generic selectors, layout-to-child misattribution, or actions visible only under certain runtime conditions), as illustrated in Fig.~\ref{fig:mot-left}.
Each candidate must therefore be refined into a locator usable by the testing engine.
Lightweight normalization can then be applied to create a PTG ready for testing.

As shown in Figure~\ref{fig:mot-left}, a Vue.js application can implement role-dependent navigation where two sidebar menu items share the same route target (``\texttt{/book}''), but appear under mutually exclusive ``\texttt{v-if}'' conditions.
In this example, the menu is rendered in a shared layout component under ``\texttt{/user}'', which provides the sidebar for its child views.
If PTG transitions are only abstracted using the route targets (i.e., collapsing actions with the same ``\texttt{from}''/``\texttt{to}''), then these two actions would be conflated into a single transition (``\texttt{/user}'' $\rightarrow$ ``\texttt{/book}''):
This could lead to infeasible or non-executable actions during testing (e.g., attempting to click a hidden menu item for the current role).
A more appropriate PTG for Vue.js should keep these as distinct edges, because they represent different \emph{executable entry points} to the same target.
In testing, conflating such actions may lead to repeated action failures (e.g., the target element not being visible under the current role), wasting time and resources, and reducing coverage.
Figure~\ref{fig:mot-right} shows each edge annotated with an executable selector (e.g., visible menu text) and condition/guard annotations that are treated as metadata.

To address these challenges to PTG-guided testing, we propose \textsc{LLMVue}, an LLM-assisted framework that generates a source-level cross-component PTG from Vue.js source code for MBT.
\textsc{LLMVue} follows a three-stage pipeline.
Stage~I statically extracts candidate edges from router configurations and SFC resources, prioritizing the avoidance of missed true transitions (high recall, i.e., retaining plausible transitions and deferring filtering to later stages).
Stage~II uses an LLM (GPT-4o~\cite{GPT4o}) to refine and enrich these candidates with executable locators and condition annotations, while preserving recall.
Stage~III applies deterministic normalization and filtering that remove structural artifacts, filter edges that mix mutually exclusive menu branches (when branch labels are available), and deduplicate redundant selectors.

The final PTG is directly used for MBT for GUIs: 
The testing engine repeatedly selects an outgoing PTG edge from the current route node and executes it using Playwright~\cite{playwright}, generating test actions.
This continues until the action budget is exhausted.
Conditions are treated as annotations, for disambiguation and debugging, and actual executability is determined at runtime, using locator visibility and interactability.
The current study focuses on systematic navigation and coverage measurement; domain-specific functional assertions are out of scope.

We used ten open-source Vue.js projects from GitHub to evaluate \textsc{LLMVue}. 
The \textsc{LLMVue} testing effectiveness was compared with random exploration, using the same action budget.
We also examined the PTG quality using precision, recall, and graph edit distance (GED)~\cite{gao2010survey}.

In summary, this paper makes the following contributions:
\begin{itemize}[leftmargin=2em, topsep=0pt, itemsep=0pt]
    \item 
    We propose \textsc{LLMVue}, an end-to-end framework for Vue.js that constructs a source-level PTG, and directly uses it to guide model-based GUI testing.
    
    \item 
    We introduce an LLM-assisted edge-refinement technique that turns (recall-focused) static candidates into executable PTG edges.      
    This is done by synthesizing executable locators and attaching condition annotations, while preserving recall.
    
    \item 
    We develop a deterministic PTG normalization process, and empirically evaluate (i) PTG modeling quality and 
    (ii) end-to-end \textsc{LLMVue} testing effectiveness.
   
\end{itemize}

The remainder of this paper is organized as follows.
Section~\ref{sec:related-work} reviews related work.
Section~\ref{sec:methodology} presents the proposed \textsc{LLMVue} framework.
Section~\ref{sec:experimental-study} describes the experimental design, and Section~\ref{sec:experimental-results} presents the results and analysis.
Section~\ref{sec:discussion} discusses the findings and implications.
Finally, Section~\ref{sec:conclusion} concludes the paper and outlines future work.

\section{Related Work} 
\label{sec:related-work}

This section briefly introduces some related work about web GUI modeling, web GUI testing, and large language models.

\subsection{Web GUI Modeling}
\label{sec:web-gui-modeling}

Web GUI modeling involves constructing abstract models that describe the navigational and structural behavior of web applications~\cite{garousi2013systematic}.
Early studies established the foundations of model inference and automated crawling for dynamic web applications and Rich Internet Applications (RIAs), including invariant-based crawling and state-space exploration techniques~\cite{mesbah2009invariant,mesbah2011invariant,mesbah2012crawling,van2015crawl,mesbah2015advances}.

\citet{dincturk2014model} presented a model-based crawling approach for Rich Internet Applications (RIAs) that constructs a state-transition model of the web interface, and applies the hypercube strategy~\cite{benjamin2011strategy} to efficiently explore dynamic web states, outperforming traditional crawling methods.
\citet{pellegrino2015jak} introduced j{\"a}k, a dynamic-analysis-based crawler that hooks client-side JavaScript APIs to construct navigation graphs for modern web applications, enabling significantly broader exploration coverage compared with traditional static URL extraction approaches.
\citet{lin2017using} developed a semantic-similarity-based approach for web-application testing that models Document Object Model (DOM) elements as vector representations.
This enables the inferencing of input topics, detection of clickable elements, and the comparison of GUI states, leading to more accurate behavior models.
\citet{ricos2025delta} proposed an approach that uses inferred state models to identify, compare, and highlight changes across desktop, web, and mobile applications:
This aims to facilitate GUI-evolution analysis and the detection of potential interface failures.

\subsection{Web GUI Testing}
\label{sec:web-gui-testing} 

Current approaches for Web GUI testing include structure-driven coverage analysis, perception-based exploration, and reinforcement learning for end-to-end interaction~\cite{li2024survey}. 
Recent surveys and benchmark studies have summarized the challenges and opportunities of automated web-agent exploration and GUI testing for modern web applications~\cite{balsam2025web}.
Model-based GUI testing uses state-transition models to systematically drive exploration and test generation~\cite{dias2007survey,belli2017advances}.

\citet{mirzaaghaei2014dom} introduced a set of DOM-based test-adequacy criteria for web applications.
They used these criteria to develop DomCovery, a tool that automatically extracts and measures coverage of DOM states, transitions, and elements.
\citet{wang2024leveraging} introduced VETL, an approach that uses large vision-language models for end-to-end automated web-GUI testing.
VETL enables context-aware text-input generation and GUI-element selection through visual question-answering:
VETL outperformed existing state-of-the-art techniques for exploration and bug-detection performance.
\citet{li2026large} investigated the effectiveness of eleven LLMs on 146 web forms from 30 open-source Java web applications.
Their study examined three HTML-structure-pruning methods to extract key contextual information for automated web-form-test generation.
\citet{gu2025deep} introduced WebRLED, an approach that uses deep reinforcement learning for automated GUI testing of complex web applications:
WebRLED uses a grid-based action-value learning technique, an action discriminator, and an adaptive curiosity-driven reward model.

Early work in model-based GUI testing focused on constructing models from functional test suites or runtime traces.
\citet{boukhris2017case} used a black-box model-based testing approach that constructs failure-mitigation models from functional test suites and uses genetic algorithms to generate and weave failure-mitigation test paths.
\citet{mattiello2022model} introduced MoLeWe, a model-based web-testing approach that infers event-driven models from existing automated GUI tests:
MoLeWe supports model reuse and the generation of new test cases.
\citet{yandrapally2022fragment} introduced FragGen, a model-based test-generation approach that uses fragment-level state abstraction to infer web-application models and generate regression test suites for dynamic web apps.
More recently, \citet{liu2026judge} proposed Judge, a model-based web GUI-testing approach that uses structure-merging and contrastive learning.

\subsection{Large Language Models for Software Testing}
\label{sec:large-language-models-software-testing}

Large Language Models (LLMs)~\cite{zhang2026instruction,zhao2023survey} have been applied to a broad range of software engineering tasks~\cite{jiang2024ledex,wang2025teaching}, including software testing~\cite{schafer2023empirical,nan2025test}.
\citet{wang2024software} reviewed 102 studies on LLMs for software testing, summarizing their use across various testing tasks, presenting commonly-adopted models and prompting techniques, and outlining key challenges and future research opportunities.
\citet{shang2025large} evaluated 37 LLMs on multiple unit-testing tasks, analyzing the effects of fine-tuning and prompt engineering, and highlighting the superior performance of fine-tuned LLMs over existing state-of-the-art approaches.
\citet{yang2024evaluation} studied LLM-based unit-test generation over 17 Java projects, evaluating five open-source LLMs, comparing them against GPT-4 and EvoSuite.
Beyond test generation, researchers have investigated LLM-assisted test execution, test maintenance, regression-test evolution, and broader evaluations of LLM-generated testing artifacts~\cite{dong2025contested,bouzenia2025you,liu2025strut,chi2025reaccept}.

\section{Methodology}
\label{sec:methodology}

This section describes our proposed end-to-end methodology for model-based GUI testing on Vue.js.
The section introduces the three-stage PTG construction pipeline: 
(1) static graph extraction, (2) LLM-based semantic refinement, and (3) graph normalization.

\subsection{Problem Formulation}
\label{sec:problem-formulation}

Let $P$ be a Vue.js project implementing a single-page application (SPA), and let $C$ be its complete source code, including the router configuration and Vue single-file components (SFCs):
Given $C$, the goal is to extract a page transition graph (PTG) that can serve as a navigation model for automated GUI testing.

A PTG is modeled as a directed labeled graph $G=(V,E)$, where $V$ is the set of route nodes and $E$ is the set of navigation edges, and each node $v\in V$ is a route path defined in the router (e.g., ``\texttt{/home}'', ``\texttt{/user/add}'').
Each edge $e\in E$ represents a navigation-relevant user interaction that transitions the application from one route to another:
$e = \langle v_{\mathrm{from}}, v_{\mathrm{to}}, \sigma, \tau, \phi \rangle$,
where $v_{\mathrm{from}}, v_{\mathrm{to}}\in V$ are the source and target routes; 
$\sigma$ is an executable selector for UI automation (Playwright locator~\cite{playwright} in this paper); 
$\tau$ is the event type (e.g., ``\texttt{click}'' or ``\texttt{redirect}''); and 
$\phi$ is an optional visibility/guard condition (e.g., extracted from ``\texttt{v-if}''/``\texttt{v-show}'').
In this paper, $\phi$ is treated as an annotation for disambiguation/debugging, and actual executability is determined at runtime by locator visibility and interactability.
The constructed PTG is denoted as $G_{\textsc{ext}}$, and the expert-constructed ground truth is $G_{\textsc{gt}}$ (Section~\ref{sec:ground_truth_construction}).

\subsection{Overview} 
\label{sec:overview}

\begin{figure*}
    \centering
    \includegraphics[width=\linewidth]{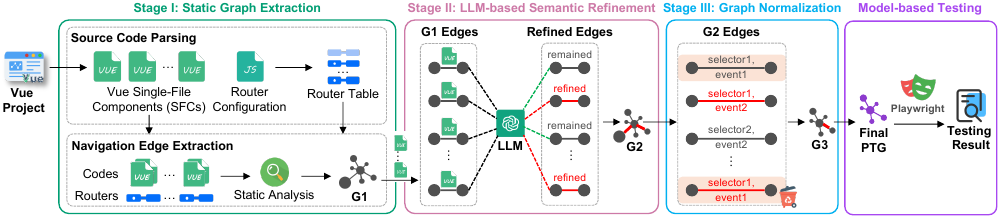}\captionsetup{skip=1pt}
    \caption{Overview of LLM-Assisted Model-Based GUI Testing for Vue.js Web Applications.}
    \label{fig:fig-overview}
\end{figure*}

As shown in Figure~\ref{fig:fig-overview}, \textsc{LLMVue} follows a three-stage pipeline for PTG construction: 
(I) Static Graph Extraction; 
(II) LLM-based Semantic Refinement; and 
(III) Graph Normalization.
During the \textbf{Static Graph Extraction}, the Vue router configuration and component templates are analyzed using Abstract Syntax Trees (ASTs), and then parsed to collect candidate route transitions.
During the \textbf{LLM-based Semantic Refinement}, 
each candidate edge is refined by enriching the UI selectors with textual cues and, when present, extracting guarding conditions:
This converts the placeholder selectors into executable Playwright locators.
Finally, the \textbf{Graph Normalization} involves removing duplicates and obvious non-transition artifacts, producing the final PTG for testing and evaluation.

The \textbf{Graph Normalization} stage produces the final PTG $G_3$ for model-based GUI testing. 
The testing engine starts from a configured entry URL and maintains the current route node. 
The engine then repeatedly selects an outgoing PTG edge and executes it with Playwright until the action budget is exhausted.

\subsection{Stage I: Static Graph Extraction}
\label{sec:stage-i-static-graph-extraction}

With a bias towards not missing true transitions (high recall), this stage produces an initial, potentially noisy, candidate PTG $G_1=(V_1, E_1)$ from the project source code. 
A main aim is to maximize coverage of candidate route-to-route transitions, including imprecise or ambiguous edges that will be refined and filtered in later stages.
In this stage, uncertain navigation targets or triggers result in multiple plausible targets being kept, and coarse, readily-instantiable selectors (e.g., tag-based, class-based, and text-based) being used as placeholders:
These selectors are not stable or test-ready, and will be converted into Playwright-ready locators in Stage~II.

\subsubsection{Source Code Parsing}

Static analysis, using ASTs, extracts the routing structure and route-level navigation information.
The \textbf{router configuration parsing} produces a router-derived \emph{route table} that contains:
(i) a set of valid route paths and their mapped page components (path$\rightarrow$component);
(ii) explicit redirection rules (from$\rightarrow$to); and
(iii) parent-child relations for nested routes.
The number of times each parent path is defined in the router configuration is calculated:
This helps manage situations in which a parent route (e.g., a layout route) is declared across multiple route objects.
Without this, child-route expansion and layout-to-child propagation may attach children to the wrong parent, producing incorrect ``\texttt{from/to}'' endpoints.
\textbf{Vue-component parsing} extracts navigation-related information from both templates and scripts.
Navigation widgets (e.g., ``\texttt{router-link}'' targets and Element UI/Plus menu items via the ``\texttt{to}''/``\texttt{index}'' attributes) and their event bindings are identified from the templates.
Router calls (e.g., {``\textttsplie{\$router.push/replace}''}) and common navigation-related code patterns (e.g., ``\texttt{window.open}'' and ``\texttt{router.resolve}'') are extracted from the scripts.

\subsubsection{Navigation Edge Extraction}

After parsing, candidate edges are constructed by matching navigation information to router-defined targets.
A component usage index maps each component file to all route paths under which it can be rendered.
This index includes both 
(i) direct router bindings, and
(ii) transitive usage inferred from a static dependency graph of component imports and template usage.

Next, two categories of edges are extracted:
(1) \textbf{Router-level edges:} 
for each explicit redirect in the route table, an edge with ``\texttt{selector="-"}'' as a placeholder selector and ``\texttt{event=redirect}'' is added.
(2) \textbf{Component-level edges:} 
for each component and associated route, the extracted navigation information (router calls, ``\texttt{router-link}'', menu items, dropdown items) is converted into candidate edges.

Navigation widgets in nested routing with a shared layout are often rendered in the layout component, with the user-visible state corresponding to the active child route.
When a layout component is associated with multiple child routes, the layout-derived menu edges are propagated to those child routes
---
this enables model executable navigation from the currently visible page.
To avoid spurious nodes, redirect-only parent routes are skipped during edge generation.

Targets are enumerated from available literals and router-aware matching: 
explicit route strings, router-link ``\texttt{to}'' values, menu ``\texttt{index}'' values, and route-table-based normalization of name-like targets to concrete paths when applicable.
Ambiguous widgets 
---
generic menu-item selectors, dynamic menu rendering, and navigation-like code such as ``\texttt{window.open}''
---
may introduce false positives in this stage:
Stage~II enriches the selectors/conditions without deleting edges, and Stage~III performs redundancy removal and filtering.

 \begin{algorithm}[!b]
    \DontPrintSemicolon
    \caption{Static Graph Extraction}
    \label{alg:stage1-extraction}
    \SetKwInOut{Input}{Input}\SetKwInOut{Output}{Output}

    \Input{Vue.js source code $\mathcal{S}$, router configuration $R$.}
    \Output{Recall-oriented candidate PTG $G_1 = (V_1, E_1)$.}

    $V_1 \leftarrow \emptyset$, $E_1 \leftarrow \emptyset$;

    $\text{routeTable} \leftarrow \texttt{\textbf{ParseRouterConfig(}}R\texttt{\textbf{)}}$;
    
    $\text{redirects} \leftarrow \text{routeTable}.\texttt{\textbf{getRedirects()}}$;
    
    $\text{componentUsageIndex} \leftarrow \texttt{\textbf{BuildComponentUsageIndex(}}\mathcal{S}, \text{routeTable}\texttt{\textbf{)}}$;

    $V_1 \leftarrow V_1 \cup \text{routeTable}.\texttt{\textbf{getAllPaths()}}$;~~~~~// Add all valid route paths (including expanded nested paths).

    \For{each entry $(\text{from}, \text{to}) \in \text{redirects}$}{
        $E_1 \leftarrow E_1 \cup \{(\text{from}, \text{to}, \text{``-''}, \text{redirect}, \text{null})\}$;
    }

    \For{each Vue file $f \in \mathcal{S}$}{
        $\text{facts} \leftarrow \texttt{\textbf{ExtractFactsFromAST(}}f\texttt{\textbf{)}}$;~~~~~// Template + script facts.
        $\text{associatedRoutes} \leftarrow \texttt{\textbf{ResolveAssociatedRoutes(}}f, \text{routeTable}, \text{componentUsageIndex}\texttt{\textbf{)}}$;

        \For{each route $r \in \text{associatedRoutes}$}{
            \If{$\texttt{\textbf{IsRedirectParentRoute(}}r, \text{routeTable}\texttt{\textbf{)}}$}{
                \textbf{continue};~~~~~// Skip redirect parent routes (heuristic).
            }
            \If{$\texttt{\textbf{ShouldSkipLayoutPropagation(}}f, r, \text{routeTable}\texttt{\textbf{)}}$}{
                \textbf{continue};
            }
            $E_1 \leftarrow E_1 \cup \texttt{\textbf{ExtractEdgesFromFacts(}}r, \text{facts}, \text{routeTable}, f\texttt{\textbf{)}}$;
        }
    }

    $E_1 \leftarrow E_1 \cup \texttt{\textbf{ExtractNestedRouteEdges(}}\text{routeTable}\texttt{\textbf{)}}$;~~~~~// Structural parent$\rightarrow$child edges from nested routing (filtered in Stage~III).

    \textbf{return} $G_1 = (V_1, E_1)$;
\end{algorithm}

Algorithm~\ref{alg:stage1-extraction} summarizes Stage~I, including router parsing, component usage resolution, and candidate edge generation with layout propagation and redirect handling.

\subsection{Stage II: LLM-based Semantic Refinement}
\label{sec:stage-ii-llm-based-semantic-refinement}

Stage~II addresses two key limitations of the static analysis:
(1) that generic selectors may lack sufficient specificity for automated testing; and 
(2) that conditional navigation semantics may not sufficiently support the lightweight static analysis.
An LLM in \emph{enhancement-only} mode enriches the candidate edge ($E_1$ in $G_1$) semantics, without deleting edges:
Stage~II preserves each edge's endpoints and event type (``\texttt{from}'', ``\texttt{to}'', ``\texttt{event}'') and only refines fields that improve executability and interpretability (mainly ``\texttt{selector}'' and ``\texttt{condition}'').
The output from Stage~II is an enhanced graph $G_2$ whose selectors are more Playwright-friendly, and whose conditions better capture when an interaction is visible/enabled.

\subsubsection{Selector Enhancement and Condition Extraction}

Given a component's source code and the extracted edges, the LLM performs four enhancement tasks.
\textbf{(1) Text content augmentation:} 
When matching text can be found in the source code, then a ``\textttsplie{:has-text(...)}'' constraint is appended to overly generic selectors (e.g., ``\texttt{button}'', ``\texttt{a}'', or bare menu-item classes).
Selectors that already have text constraints are left unchanged.
\textbf{(2) Framework-specific normalization:} 
The selectors are normalized to reflect the rendered DOM under different UI libraries and Vue versions:
Typical rendered elements/classes include
``\texttt{<el-button>}'' rendered as a ``\texttt{button}'' element, and ``\texttt{<el-dropdown-item>}'' as a ``\textttsplie{.el-dropdown-menu\_\_item}''.
The submenu class naming is also adjusted from  Element UI (``\textttsplie{.el-submenu}'') and Element Plus (``\texttt{.el-sub-menu}'').
\textbf{(3) Condition extraction:} 
The \emph{visibility/enabling} conditions that control whether an interaction is present or actionable are then extracted.
These conditions can come from template directives (e.g., ``\textttsplie{v-if/v-show}''), or from script-level conditional menu construction.
When an edge connects pages that appear in different conditional branches, the relation is annotated in the \texttt{condition} field using ``\texttt{CROSS\_ROLE: <from\_cond> -> <to\_cond>}''
---
``\texttt{CROSS\_ROLE}'' is inherited from role-dependent menu scenarios, but it generally denotes any mutually exclusive branch condition (e.g., feature flags).
``\texttt{CROSS\_ROLE}'' is treated as an annotation string in downstream processing, rather than a boolean expression to be evaluated.
\textbf{(4) Recall preservation with format validation:} 
The LLM is constrained to return all input edges (no deletions). 
The output is checked (validated) to ensure that it follows the edge schema and that each route target is valid (according to the route table):
If validation fails after retries, the original Stage~I edges for that file are retrieved.

\begin{algorithm}[!t]
    \DontPrintSemicolon
    \caption{LLM-based Semantic Refinement}
    \label{alg:stage2-refinement}
    \SetKwInOut{Input}{Input}\SetKwInOut{Output}{Output}

    \Input{Initial graph $G_1 = (V_1, E_1)$, Vue source files $\mathcal{S}$, LLM client $\mathcal{L}$.}
    \Output{Enhanced graph $G_2 = (V_2, E_2)$.}

    $V_2 \leftarrow V_1$, $E_2 \leftarrow \emptyset$;

    $\text{edgesByFile} \leftarrow \texttt{\textbf{GroupEdgesBySourceFile(}}E_1\texttt{\textbf{)}}$;~~~~~// By each edge's source component file.

    \For{each file $f \in \text{edgesByFile.keys()}$}{
        $\text{edges} \leftarrow \text{edgesByFile}[f]$;

        $\text{sourceCode} \leftarrow \texttt{\textbf{ReadFile(}}f\texttt{\textbf{)}}$;

        \If{$\texttt{\textbf{IsRouterConfigFile(}}f\texttt{\textbf{)}}$}{
            $E_2 \leftarrow E_2 \cup \text{edges}$;~~~~~// Router edges are programmatic; no selector/condition enrichment needed.

            \textbf{continue};
        }

        $\text{batches} \leftarrow \texttt{\textbf{SplitIntoBatches(}}\text{edges}, 30\texttt{\textbf{)}}$;~~~~~// Token-budget driven batching.

        \For{each batch $B \in \text{batches}$}{
            $\text{validatedEdges} \leftarrow \emptyset$;

            \For{$\text{attempt} = 1$ \KwTo $3$}{~~~~~// Retries for occasional JSON/field errors.
            
                $\text{systemPrompt} \leftarrow \texttt{\textbf{BuildSystemPrompt()}}$;~~~~~// Enhancement rules.

                $\text{userPrompt} \leftarrow \texttt{\textbf{BuildUserPrompt(}}B, \text{sourceCode}\texttt{\textbf{)}}$;

                $\text{response} \leftarrow \mathcal{L}.\texttt{\textbf{ask(}}\text{systemPrompt}, \text{userPrompt}\texttt{\textbf{)}}$;

                $\text{parsedEdges} \leftarrow \texttt{\textbf{ParseJSON(}}\text{response}\texttt{\textbf{)}}$;

                \If{$\texttt{\textbf{ValidateResponse(}}\text{parsedEdges}, B\texttt{\textbf{)}}$}{
                    $\text{validatedEdges} \leftarrow \text{parsedEdges}$;

                    \textbf{break};
                }
                \Else{
                    $\text{userPrompt} \leftarrow \texttt{\textbf{BuildReflectionPrompt(}}\text{userPrompt}, \text{response}\texttt{\textbf{)}}$;
                }
            }

            \If{$\text{validatedEdges} = \emptyset$}{
                $\text{validatedEdges} \leftarrow B$;~~~~~// Fallback to original edges.
            }

            $E_2 \leftarrow E_2 \cup \text{validatedEdges}$;
        }
    }

    \textbf{return} $G_2 = (V_2, E_2)$;
\end{algorithm}

Algorithm~\ref{alg:stage2-refinement} comprises three main steps:
In the \textbf{File Grouping}, the edges are first grouped by their source file as the full context to the LLM.
During the \textbf{Iterative Refinement}, each LLM call is retried up to three times (if validation fails), using feedback-based re-prompting.
Finally, if all attempts fail, then the \textbf{Fallback Mechanism} returns the original edges, to ensure no data loss.

\subsubsection{Prompt Engineering and Validation}

The LLM interaction employs a two-part prompt design.
The \textbf{system prompt} enforces enhancement-only mode, preserves selector kind with optional text constraints, applies project-specific normalization (e.g., UI library differences), and extracts conditions under a fixed schema; router-config edges with placeholder selectors are excluded from modification.
The \textbf{user prompt} provides the source file content and a JSON list of edges.
If files exhibit conditional menu rendering, then the full file is included to enable reasoning over script-level branches.
Otherwise, only template-related content is included.
The decision is guided by Stage~I navigation information: 
If script-level menu or branching information is detected for a file, then the script is included to avoid missing branch-dependent conditions.

Response validation checks three criteria:
(1) that the edge count matches the input;
(2) that fields are present (``\texttt{from}'', ``\texttt{to}'', ``\textttsplie{selector}''); and 
(3) that each target route is a valid router path, after route-table normalization.
For example, an unambiguous base path (e.g., ``\texttt{/create}'') is normalized to a parameterized route (e.g., ``\textttsplie{/create/:id}''); otherwise, the original target is kept.
If the validation fails, \textbf{reflection prompting} appends the previous response and validation errors to the user prompt, and requests corrections.
This iterative refinement can be attempted up to three times: 
If all three attempts fail, the system falls back to the original Stage~I edges for that file.
This fallback aims to avoid introducing new noise: 
No new edges are created, and the original edges are preserved.

\subsection{Stage III: Graph Normalization}
\label{sec:stage-iii-graph-normalization}

Stage~III performs \emph{edge-level} normalization on $G_2$ to produce the final PTG, $G_3$.
Stage~III removes structural edges and global redirect artifacts, filters cross-group edges derived from script-level conditional menus, removes most self-loops (edges whose endpoints are the same), and deduplicates redundant selectors.

\subsubsection{Redundancy Removal}
Lightweight normalization and filtering rules are used to reduce the obvious noise introduced by high-recall extraction and LLM enhancement.
During the \textbf{selector and condition post-processing}, string-level normalization is applied
---
no edges are deleted in this step.
For submenu selectors (e.g., \texttt{.el-submenu}) lacking a text anchor, the title text is extracted from the source file (if possible), and a \texttt{:has-text(...)} constraint is added:
If the extraction fails, then the original selector is kept unchanged.
The synthetic prefix \texttt{router-link:} is converted to \texttt{a:}, as a \emph{heuristic} for the default Vue Router rendering.
Conditional expressions are normalized and converted into standard notations.
During the \textbf{(structural edge filtering},
the router-configuration structure edges with \texttt{event=route} are dropped
---
these represent parent-child relations in the route table, not a user-triggered interaction.
The wildcard auto-redirect artifacts with \texttt{from=/*} and \texttt{selector="-"} are also dropped
---
they represent global redirect behavior, not local page UI navigation.
During the \textbf{conditional-branch consistency filtering}, a lightweight \emph{branch label} (page-group id) is assigned to each route path referenced in a script-level if/else menu branch.
The branch label is a coarse, file-local
syntactic identifier of the menu-construction branch.
Edges are only filtered when \emph{both} endpoints are labeled, and the labels disagree.
Routes without a branch label 
---
due, perhaps, to non-literal route construction or menu generation through helper functions
---
are treated as unlabeled, and are not filtered by this rule.
\textbf{Self-loop filtering:} 
Because the PTG primarily models route-to-route navigation, self-loops are treated as auxiliary UI-structure actions.
Most self-loops are removed, with only a few surviving a conservative rule-based filter:
These self-loops represent the UI expansion steps (e.g., opening a dropdown/submenu) required to reach nested actionable items.

\subsubsection{Graph Deduplication}
Two edges are considered duplicates if they have the same endpoints and interaction signature 
--- 
the same from, to, event, and selector;
the \texttt{condition} is not a deduplication key.
Multi-stage grouping is used to select representatives.
When multiple edges share the same key (e.g., differing only in \texttt{condition}), then only the first-seen edge is kept.

The deduplication process operates at three levels, and is implemented as a set of practical, selector-driven heuristics.
\textbf{(1) Selector-level deduplication:} Edges are grouped by \texttt{(from, to, event)}, and exact duplicate selectors within each group are removed.
When there are both generic and more specific selectors for the same interaction,
the selectors that contain a class/id or an explicit text constraint are kept.
\textbf{(2) Route-pair preference filtering:} Edges are grouped by \texttt{(from, to)}, and the more stable and actionable selectors are prioritized for keeping. 
Multiple edges may be retained for a route pair when they are of different actionable kinds.
\textbf{(3) Submenu-prefix deduplication:} For Duplicate Element UI/Plus selectors, that differ \emph{only} by a submenu prefix (e.g., \texttt{.el-submenu .el-menu-item} vs. \texttt{.el-menu-item}) are identified (by grouping on \texttt{(from, to, extractedText, baseSelector)}).
For cross-page navigation, versions \emph{without} the prefix (less sensitive to submenu expansion state) are preferred. 
for self-loops, the versions \emph{with} the prefix (more specific) are preferred.
Deduplication keys are selector-driven:
If multiple edges differ only in the \texttt{condition} field, but share the same endpoints/event/selector, then only a single representative edge is kept.

\begin{algorithm}
    \DontPrintSemicolon
    \caption{Graph Normalization}
    \label{alg:stage3-normalization}
    \SetKwInOut{Input}{Input}\SetKwInOut{Output}{Output}

    \Input{Enhanced graph $G_2 = (V_2, E_2)$.}
    \Output{Final graph $G_3 = (V_3, E_3)$.}

     $V_3 \leftarrow V_2$;
     $E_3 \leftarrow E_2$;

     $E_3 \leftarrow \texttt{\textbf{PostProcessSelectorsAndConditions(}}E_3\texttt{\textbf{)}}$;

     $E_3 \leftarrow \texttt{\textbf{SelectorLevelDedup(}}E_3\texttt{\textbf{)}}$;~~~~~// Group by (from,to,event).

     $E_3 \leftarrow \texttt{\textbf{RoutePairPreferenceFilter(}}E_3\texttt{\textbf{)}}$;~~~~~// Group by (from,to).

     $E_3 \leftarrow \texttt{\textbf{FilterStructuralEdges(}}E_3\texttt{\textbf{)}}$;~~~~~// Drop event=route and wildcard auto-redirect artifacts.

     $E_3 \leftarrow \texttt{\textbf{DedupSubmenuPrefix(}}E_3\texttt{\textbf{)}}$;~~~~~// Element UI/Plus submenu-prefix duplicates.

     $E_3 \leftarrow \texttt{\textbf{FilterCrossGroupEdges(}}E_3\texttt{\textbf{)}}$;~~~~~// Conditional menu branch consistency.

     $E_3 \leftarrow \texttt{\textbf{FilterSelfLoops(}}E_3\texttt{\textbf{)}}$;~~~~~// Conservative rule-based self-loop filtering.

     \textbf{return} $G_3 = (V_3, E_3)$;
\end{algorithm}

Algorithm~\ref{alg:stage3-normalization} presents Stage~III as a sequence of post-processing, filtering, and deduplication steps.

\section{Experimental Study}
\label{sec:experimental-study}

This section presents details of the experimental study, including the research questions, datasets, evaluation metrics, and implementation. 

\subsection{Research Questions} 
\label{sec:research-questions}

This research aimed to evaluate \textsc{LLMVue}, an LLM-assisted approach that constructs PTGs from Vue.js source code and uses them to drive model-based GUI testing.
The following two research questions guided the study:
\begin{description}[leftmargin=2em, topsep=0pt, itemsep=0pt]
    \item 
    \textbf{RQ1} (\textbf{Quality of \textsc{LLMVue}-Constructed PTGs}):
    What is the quality of the PTGs constructed by \textsc{LLMVue} for Vue.js applications?

    \item 
    \textbf{RQ2} (\textbf{Impact of Using \textsc{LLMVue}-Constructed PTGs on Automated Testing}):
    What is the impact of using \textsc{LLMVue}-constructed PTGs on automated GUI testing?
\end{description}

\textbf{RQ1} was addressed by comparing the PTGs constructed by \textsc{LLMVue} against manually-constructed ground-truth graphs, using precision, recall, F1-score, and graph edit distance (GED).
To answer \textbf{RQ2}, testing effectiveness was evaluated under the same action budget by measuring statement and page coverage during automated GUI exploration, and comparing \textsc{LLMVue}-guided testing with budget-matched random exploration.
RQ2 is positioned as a focused case study on two representative subjects and is intended to provide preliminary evidence rather than to offer a general comparison with state-of-the-art GUI testing techniques.

\subsection{Datasets}
\label{sec:datasets}

Following prior empirical studies that curate open-source web applications from GitHub (e.g., \citet{li2026large}), we constructed the dataset by querying GitHub with Vue.js SPA-related keywords. Using GitHub's \textit{Best match} option, we initially retrieved approximately 6.1k repositories and then manually screened them according to our inclusion criteria. 
We included projects that (1) are Vue.js SPAs with Vue Router; (2) provide non-trivial route-level navigation; and (3) can be built and executed end-to-end under version-locked snapshots so that Playwright-based exploration is reproducible.
We excluded demo-only projects, projects without runnable instructions, and projects with missing dependencies or back-end services that could not be reliably deployed.
The final dataset size (ten projects) is consistent with sample sizes commonly used in prior model-based GUI testing studies for HarmonyOS apps \cite{chen2024model}.

 \begin{table}[!b]
    \centering
    \caption{Details of the ten Vue projects used in the experiments.}    
    \label{tab:dataset-summary}
    \setlength{\tabcolsep}{4mm}
    \begin{tabular}{llrrrl}
        \hline
        \textbf{ID} & \textbf{Project Name} & \textbf{\#SFC} & \textbf{LOC} & \textbf{\#Routes} \\
        \hline
        P1  & Vue-Springboot-Library      &14  &2,459  &10 \\
        P2  & questionnaire               &16  &2,612  &7  \\
        P3  & dormitory\_springboot       &23  &2,781  &24  \\
        P4  & music-website               &15  &3,227  &10  \\
        P5  & Blog                        &17  &3,480  &18  \\
        P6  & COVID                       &11  &2,252  &9  \\
        P7  & VehicleMangement-Web        &13  &3,035  &13  \\
        P8  & UniTakeout                  &10  &3,223  &8  \\
        P9  & stu\_exam\_sys              &23  &10,184  &21  \\
        P10 & house\_rental\_management   &29  &4,941  &13  \\
        \hline
    \end{tabular}
\end{table}

For reproducibility, the repositories were mirrored locally, and all experiments were conducted on these fixed snapshots.
The repository URLs, commit hashes, and build/run commands are included in our replication package~\cite{replicationpackage}.
Table~\ref{tab:dataset-summary} summarizes the dataset details:
\#SFC denotes the number of Vue single-file components (``\texttt{.vue}'');
LOC denotes the number of frontend source lines of code (excluding dependencies); and 
\#Routes denotes the number of distinct route paths in the router configuration.

\subsection{Ground Truth Construction} 
\label{sec:ground_truth_construction}

To support fair and reproducible evaluation of the LLM-constructed navigation graphs, 
the ground-truth (GT) SPA navigation graphs were constructed using a rule-based expert protocol.
This protocol was co-developed with Vue/Element UI practitioners (e.g., developers with 5+ years of Vue.js experience) and \emph{version-locked before annotation} (no rule changes during labeling).
Experts followed deterministic rules for node/edge inclusion, navigation information prioritization, selector normalization, and ambiguity resolution.

\subsubsection{Navigation Graph Model}
SPA navigation was modeled as a directed labeled graph $G=(V,E)$, where $V$ and $E$ denote route nodes and navigation edges, respectively.
\textbf{Nodes} are normalized route paths derived from router configurations and explicit router targets in scripts, with deduplication applied.
\textbf{Edges} capture navigation-relevant interactions with attributes ``\texttt{\{from, to, selector, event, selectorKind, condition, note\}}'', where \texttt{selector} is a Playwright selector, and the \texttt{condition/note} stores dynamic parameters or other metadata.
The selectorKind distinguishes extraction sources: ``\texttt{ROUTER}'' (route config), ``\texttt{SCRIPT}'' (router calls), ``\texttt{INDEX}'' (inherent menu navigation), and ``\texttt{TEXT}'' (fallback by visible text).

\subsubsection{Information Sources and Priority Rules}
GT extraction uses the following information sources and priorities:
The \textbf{route table} defines the global node set and redirects; 
each redirect yields a ``\texttt{ROUTER}'' edge with ``\texttt{event=redirect}'' and ``\texttt{note=REDIRECT}''.
\textbf{Scripts} provide navigation information through ``\textttsplie{\$router.push/replace/go}'' calls, including those inside nested callbacks.
\textbf{Templates} yield ``\texttt{INDEX}'' edges from router-enabled navigation widgets; and 
event bindings must be resolved to script handlers.
\textbf{Conflict rule:} 
If an element has both inherent navigation and script routing, GT prioritizes the ``\texttt{SCRIPT}''.

\subsubsection{Edge Definition and Inclusion/Exclusion Rules}
The following ground-truth labeling rules specify which UI interactions constitute PTG edges and how to normalize edge descriptions to ensure consistent, reproducible annotation.
\textbf{Include:} 
(i) script-based routing; 
(ii) router-enabled menu navigation; 
(iii) post-submit routing in handlers; 
(iv) submenu expand/collapse as explicit self-loops.
\textbf{Exclude:} 
Actions without route changes, including query/reset/save operations, dialog open/close, clipboard/share, pagination refresh on the same route, and API-only operations.
\textbf{Selector normalization:} 
Use stable rendered DOM classes with ``\texttt{:has-text}'' constraints; 
add parent context for nested menus;
scope dialog actions by dialog container; and prohibit container-only selectors.
\textbf{Dropdown rule:} 
Only dropdown items that lead to routing are edges; opening the dropdown trigger is not an edge.
\textbf{Redirect-parent propagation:} 
For a parent route redirected to target $T$, DOM-derived edges from the parent component are generated only for $T$.
\textbf{Handler tracing:} 
Trace template handlers to methods and include router calls in nested callbacks; 
for dynamic targets, record the base path and store parameters in \texttt{condition}.
\textbf{Self-loop constraint:} 
Allow self-loops only for navigation-structure components; filter spurious button self-loops, unless routing semantics exist.

\subsubsection{Consistency Controls}
The annotator variance is reduced through: 
(i) pre-annotation calibration, using shared examples under the version-locked rules;
(ii) centralized arbitration for ambiguous cases; and 
(iii) automated sanity checks (using selector format, legal \texttt{selectorKind}, route existence, and self-loop legality).

\subsection{Evaluation Metrics} 
\label{sec:evaluation-metrics}

Following previous studies~\cite{chen2024model}, two types of metrics are used in the evaluations: 
(1) \textit{graph-based metrics}, which evaluate the structural quality and similarity of the generated models; and 
(2) \textit{testing-based metrics}, which measure the effectiveness of automated GUI testing guided by these models.

\subsubsection{Graph-Based Metrics}
Four widely-used metrics were adopted to evaluate the structural quality of the constructed graphs:
\begin{itemize}
    \item \textbf{Precision} measures the ratio of correctly identified edges to all detected edges:
            \begin{equation}
                \text{Precision} = \frac{TP}{TP + FP},
            \end{equation}
            where $TP$ (true positives) denotes the number of edges that are correctly detected and exist in the Vue.js application, and $FP$ (false positives) denotes the number of edges that are detected by the model, but do not actually exist in the Vue.js application.
    \item \textbf{Recall} measures the ratio of correctly detected edges to all actual edges: 
            \begin{equation}
                \text{Recall} = \frac{TP}{TP + FN},
            \end{equation}
            where $FN$ (false negatives) denotes the number of actual edges that exist in the Vue.js application but that are missing from the generated graph;

    \item \textbf{F1-Score} is the harmonic mean of precision and recall:
            \begin{equation}
                \text{F1} = \frac{2 \times \text{Precision} \times \text{Recall}}{\text{Precision} + \text{Recall}}.
            \end{equation}

    \item \textbf{Graph Edit Distance (GED)} measures the minimum number of edit operations required to transform the constructed graph $G_p$ into the ground-truth graph $G_t$:
            \begin{equation}
                \text{GED} (G_p, G_t) = \min_{\text{edit seq.}} |\text{operations}|.
            \end{equation}
            GED is a structural similarity metric that measures the minimum number of edit operations (insertion, deletion, or substitution of edges) required to transform the LLM-generated graph $G_p$ into the manually constructed ground-truth graph $G_t$.
\end{itemize}

\subsubsection{Testing-Based Metrics}
Three runtime metrics were used to evaluate the testing effectiveness of the automated GUI testing based on generated PTGs.
\begin{itemize}
    \item \textbf{Action Number (AN)} counts the total number of user actions (e.g., clicks, inputs, and navigations) executed within a fixed time budget.

    \item \textbf{Statement Coverage (SC)} measures the ratio of executed JavaScript statements to total statements --- this was collected using the \textit{nyc}~\cite{nyc} tool.

    \item \textbf{Page Coverage (PC)} is the number of distinct Vue pages visited during testing as a proportion of the total number of pages defined in the routing configuration.
\end{itemize}

\subsection{Implementation Details} 
\label{sec:running-environment}

All experiments were conducted on a MacBook Pro laptop with an Apple M3 Max processor and 64GB of RAM.
The Vue PTG extraction module was implemented in Java using JDK 17.
The graph-based experimental data collection was conducted using Node.js:
The Node.js runtime version was v18.20.8, and the corresponding npm version was 10.8.2.
Google Chrome version 143.0.7499.194 (official version) was used.

\section{Experimental Results} 
\label{sec:experimental-results}

This section presents the experimental results and answers the two RQs from Section~\ref{sec:research-questions}.

\subsection{RQ1: Quality of \textsc{LLMVue}-Constructed PTGs} 
\label{sec:res-rq1}

This section evaluates the quality of the \textsc{LLMVue}-constructed PTGs.
The PTG $G_{\textsc{ext}}$ produced by \textsc{LLMVue} is compared against the expert-constructed ground truth $G_{\textsc{gt}}$ (Section~\ref{sec:ground_truth_construction}).
The precision, recall, and F1 are computed at the \emph{edge level}, quantifying correctness and completeness of predicted transitions.
The GED captures the overall structural distance between $G_{\textsc{ext}}$ and $G_{\textsc{gt}}$.
Each extraction was run three times per project, using deterministic decoding (GPT-4o with temperature = 0).

\begin{table}[b]
    \centering
    \caption{Quality of PTGs extracted by \textsc{LLMVue} on the ten Vue projects}    
    \label{tab:rq1-results}
    \setlength{\tabcolsep}{2.2mm}
    \begin{tabular}{lrrrrr}
        \hline
        \textbf{Project ID} & \textbf{Precision (\%) ($\uparrow$)} & \textbf{Recall (\%) ($\uparrow$)} & \textbf{F1-Score (\%) ($\uparrow$)} & \textbf{GED ($\downarrow$)} & \textbf{F1 Range}\\
        \midrule
        P1 & 98.65 & 96.05 & 97.33 & 4 & $\ge 97\%$ \\
        P2 & 100.00 & 93.33 & 96.55 & 2 & $[95\%,~97\%)$\\
        P3 & 100.00 & 94.04 & 96.93 & 13 & $[95\%,~97\%)$\\
        P4 & 93.48 & 84.31 & 88.66 & 11 & $<90\%$\\
        P5 & 90.80 & 84.95 & 87.78 & 44 & $<90\%$\\
        P6 & 100.00 & 80.00 & 88.89 & 2 & $<90\%$\\
        P7 & 91.18 & 82.30 & 86.51 & 29 & $<90\%$\\
        P8 & 95.00 & 90.48 & 92.68 & 6 & $[90\%,~95\%)$\\
        P9 & 100.00 & 82.47 & 90.39 & 27 & $[90\%,~95\%)$\\
        P10 & 76.60 & 100.00 & 86.75 & 22 & $<90\%$\\
        \hline
        Avg. &94.57  &88.79  &91.25  &16 & -- \\
        \hline
    \end{tabular}
\end{table}

Table~\ref{tab:rq1-results} presents the PTG quality details across the ten Vue projects.
The table includes the F1 range for each project, indicating whether mismatches are balanced or one-sided (high precision/low recall or vice versa).

\subsubsection{Observations}
Based on the experimental data, we have the following observations:
\begin{itemize}[leftmargin=2em, topsep=0pt, itemsep=0pt, parsep=0pt, partopsep=0pt]
    \item 
    \textbf{\textsc{LLMVue} achieves strong overall extraction quality on the benchmark set.}
    As shown in Table~\ref{tab:rq1-results}, across the ten projects, \textsc{LLMVue} achieves an average precision of 94.57\%, recall of 88.79\%, and F1-score of 91.25\%.
    The performance is also consistently high at the distribution level: 
    5/10 projects are in F1\,$\geq$\,90\% (P1, P2, P3, P8, and P9); and 
    3/10 projects exceed 95\% (P1, P2, and P3).
    The F1-score range is 86.51\%--97.33\%, with a relatively tight dispersion ($std$, the standard deviation, $\approx 4.09$), compared to precision/recall:
    This indicates that the overall balance is relatively stable across projects.
    The concentration is also reflected in Table~\ref{tab:rq1-results}: 
    Multiple projects lie on or above the 90\% F1 range, with those falling below it mainly due to asymmetric mismatches (high precision but lower recall, or vice versa).
    Overall, these results indicate that \textsc{LLMVue} recovers a large proportion of ground-truth transitions with relatively few spurious edges, on most projects.
    
    \item 
    \textbf{Precision is typically high, while recall varies with condition-dependent navigation.}
    Precision is generally high across the projects (76.60\%--100.00\%, median $\approx$ 96.83\%), and four projects (P2, P3, P6, and P9) have perfect precision (100\%).
    This suggests that once \textsc{LLMVue} determines that a transition exists, the semantic refinement and normalization steps effectively suppress spurious edges and produce correct endpoint mappings.
    In contrast, the recall has a wider spread (80.00\%--100.00\%, median $\approx$ 87.72\%).
    Table~\ref{tab:rq1-results} makes this asymmetry explicit:
    P6 and P9, for example, have perfect precision, but substantially lower recall (P6: 100.00\% vs.\ 80.00\%; P9: 100.00\% vs.\ 82.47\%).
    These missed edges reduce the recall without necessarily introducing incorrect edges, thus preserving high precision.
    
    \item 
    \textbf{GED reflects the concentration of mismatches in a small set of difficult edge patterns.}
    The mean GED is 16, but the distribution is skewed ($\text{min}=2$, $\text{median}=12$, $\text{max}=44$; and $\text{std} \approx 13.34$):
    This indicates that most projects are structurally close to the ground truth, but that a minority contribute disproportionately to the average distance.
    Specifically, four projects have very small edit distances ($\text{GED} \leq 6$ for P1, P2, P6, and P8):
    This suggests that their constructed graphs require very few edit operations to match $G_{\textsc{gt}}$.
    In contrast, several projects have markedly larger distances (e.g., P5: 44; P7: 29; P9: 27; and P10: 22), highlighting more extensive mismatches.
    Because GED aggregates different mismatch types, it should be interpreted together with precision/recall:
    (i) \emph{under-approximation} (many false negatives) increases GED because converting $G_{\textsc{ext}}$ to $G_{\textsc{gt}}$ may require the insertion of missing ground-truth edges
    ---
    this typically co-occurs with lower recall (e.g., P6 and P9); and
    (ii) \emph{over-approximation} (many false positives) increases GED by requiring deletions
    ---
    this often co-occurs with lower precision (e.g., P10).
    This connection is consistent with the data in Table~\ref{tab:rq1-results}: 
    Projects in the lower F1 bands tend to exhibit one-sided mismatches (high precision/low recall or high recall/low precision), rather than degrading uniformly along both axes.
\end{itemize}

\subsubsection{Mismatch Analysis}

To make the results in Table~\ref{tab:rq1-results} actionable, a mismatch-driven analysis was performed between the constructed PTG $G_{\textsc{ext}}$ and the ground-truth PTG $G_{\textsc{gt}}$.
First, the mismatched edge sets were computed, including both the false negatives (edges present in $G_{\textsc{gt}}$, but missing from $G_{\textsc{ext}}$) and the false positives (edges in $G_{\textsc{ext}}$ that were not in $G_{\textsc{gt}}$).
These mismatches were traced back to the corresponding Vue.js project source code to identify the underlying causes:
They were then grouped into recurring patterns.
The two main mismatch sources were:

(1)~\textbf{False negatives are mainly caused by conditional and callback-driven routing.}
A substantial proportion of the false negatives correspond to transitions that are feasible only under specific runtime states (e.g., validation succeeds, an asynchronous request returns success) or that are executed inside nested callbacks (e.g., confirmation dialogs).
From a static analysis perspective, these edges require simultaneously reasoning about 
(i) event-to-handler bindings, 
(ii) control-flow conditions, and 
(iii) asynchronous callback chains.
These transitions may be under-approximated when the routing call is not directly adjacent to the triggering widget in the source code.

(2)~\textbf{False positives are mainly caused by menu selector ambiguity and cross-role navigation noise.}
False positives primarily arise when a UI selector is not sufficiently discriminative (e.g., a generic menu-item selector without text constraints):
This leads to multiple menu entries being conflated during extraction.
Applications with role-dependent menus may also expose cross-role entry points or ``\texttt{home/index}'' shortcuts that are clickable in the UI, but are excluded by the ground-truth protocol (when they do not represent stable route-level navigation under a fixed role context).
These cases reduce precision by introducing edges that are syntactically plausible, but semantically inconsistent with the evaluation scope.

\finding{1}{\textsc{LLMVue} extracted high-quality PTGs from the selected ten Vue.js projects, achieving an average F1-score of 91.25\%.
Five projects achieve $\textrm{F1} \ge 90\%$, including three with $\textrm{F1} \ge 95\%$.
Overall, the precision was consistently high, with mismatches mainly due to missed edges (lower recall) on some projects.}

\subsection{RQ2: Impact of Using \textsc{LLMVue}-Constructed PTGs on Automated Testing} 
\label{sec:res-rq2}

This section compares the \textsc{LLMVue}-guided exploration (using the PTG constructed by \textsc{LLMVue}) with a random exploration baseline, under the same time budgets.

\begin{table}[!b]
    \centering
    \caption{Comparison of the testing effectiveness of \textsc{LLMVue}-guided vs. random exploration. 
    The 95\% CIs are nonparametric bootstrap CIs of the mean (5000 resamples; seed=42). 
    For $\Delta$PC and $\Delta$SC, pp denotes percentage points.}
    \label{tab:rq2-timebudget}
    \setlength{\tabcolsep}{0.8mm}
    \begin{tabular}{@{}llrrrrrrr@{}}
        \hline
        \textbf{ID} & \textbf{Budget} & \textbf{Mode} & \textbf{AN} & \textbf{PC (\%)} & \textbf{SC (\%)} & \textbf{$\Delta$AN} & \textbf{$\Delta$PC} & \textbf{$\Delta$SC} \\
        \hline
        \multirow{6}{*}{P1}
        & \multirow{2}{*}{60s} & LLMVue & $95.8\pm24.9$ & $67.00\pm4.50$ & $31.44\pm1.64$ & \multirow{2}*{$35.7 (59.4\%)$} & \multirow{2}*{$+13.00$\,pp} & \multirow{2}*{$+6.34$\,pp} \\
        &                      & Random & $60.1\pm24.3$ & $54.00\pm10.00$ & $25.11\pm3.11$  \\\cline{2-9}
        & \multirow{2}{*}{300s} & LLMVue & $518.0\pm32.2$ & $70.00\pm0.00$ & $38.89\pm1.12$ & \multirow{2}*{$207.4 (66.8\%)$} & \multirow{2}*{$+3.00$\,pp} & \multirow{2}*{$+7.52$\,pp} \\
        &                      & Random & $310.6\pm186.7$ & $67.00\pm3.50$ & $31.37\pm2.15$ \\\cline{2-9}
        & \multirow{2}{*}{600s} & LLMVue & $1038.3\pm35.1$ & $70.00\pm0.00$ & $41.63\pm0.67$ & \multirow{2}*{$379.5 (57.6\%)$} & \multirow{2}*{$+0.00$\,pp} & \multirow{2}*{$+7.16$\,pp} \\
        &                      & Random & $658.8\pm385.8$ & $70.00\pm0.00$ & $34.47\pm0.93$ \\
        \hline
        \multirow{6}{*}{P3}
        & \multirow{2}{*}{60s} & LLMVue & $182.4\pm26.1$ & $58.33\pm 1.67$ & $23.99\pm2.58$ & \multirow{2}*{$26.2 (16.8\%)$} & \multirow{2}*{$+5.42$\,pp} & \multirow{2}*{$-3.81$\,pp} \\
        &                      & Random & $156.2\pm40.5$ & $52.92\pm6.25$ & $27.80\pm3.12$ \\\cline{2-9}
        & \multirow{2}{*}{300s} & LLMVue & $1011.8\pm63.7$ & $66.67\pm2.50$ & $37.93\pm1.39$ & \multirow{2}*{$405.3 (66.8\%)$} & \multirow{2}*{$+5.83$\,pp} & \multirow{2}*{$-0.51$\,pp} \\
        &                      & Random & $606.5\pm92.3$ & $60.83\pm1.87$ & $38.44\pm0.91$ \\\cline{2-9}
        & \multirow{2}{*}{600s} & LLMVue & $1911.1\pm87.7$ & $67.50\pm2.29$ & $44.32\pm1.38$ & \multirow{2}*{$752.4 (64.9\%)$} & \multirow{2}*{$+5.83$\,pp} & \multirow{2}*{$+4.17$\,pp} \\
        &                      & Random & $1158.7\pm54.4$ & $61.67\pm 1.88$ & $40.15\pm0.82$ \\
        \hline
    \end{tabular}
\end{table}

\subsubsection{Project Selection and Experimental Settings}
Following the experimental design principle used in other MBT work~\cite{chen2024model}, the project with the most edges was selected to represent structurally complex applications.
P3 was selected because \textsc{LLMVue} constructed the largest PTG for it.
P1 was also included, because it had the highest extraction quality for RQ1 ($\text{F1}=97.33\%$, in Table~\ref{tab:rq1-results}).
This complementary selection aimed to reduce potential project-selection bias and to assess testing impact under both 
(i) a structurally complex project, and 
(ii) a high-quality constructed model.
For each project, we compare \textsc{LLMVue}-guided exploration against a random exploration baseline across 10 time budgets (60s-600s, $\text{step}=60s$).
Each configuration is executed 10 times.
We use random exploration as a lower-bound baseline to assess the benefit of PTG guidance under identical budgets. 
Other GUI-testing tools are excluded because a fair comparison in our setup is non-trivial (e.g., toolchain mismatches and varying crawlability)~\cite{alshahwan2012crawlability}.

\begin{figure}[t]
    \centering
    \begin{subfigure}[t]{0.325\linewidth}
        \centering
        \includegraphics[width=\linewidth]{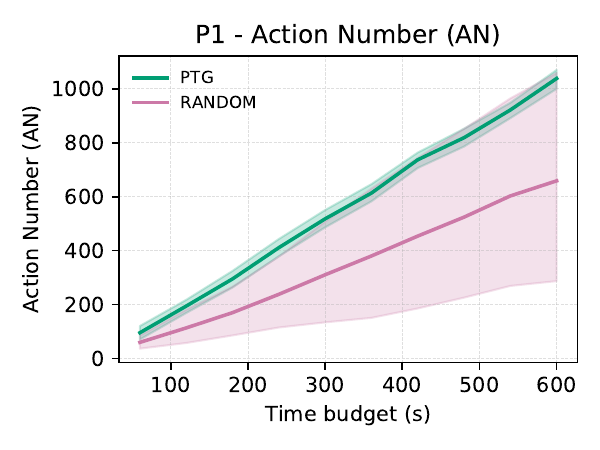}
        \captionsetup{skip=0pt}
        \caption{P1}
        \label{fig:rq2-p1-mean-an}
    \end{subfigure}
    \begin{subfigure}[t]{0.325\linewidth}
        \centering
        \includegraphics[width=\linewidth]{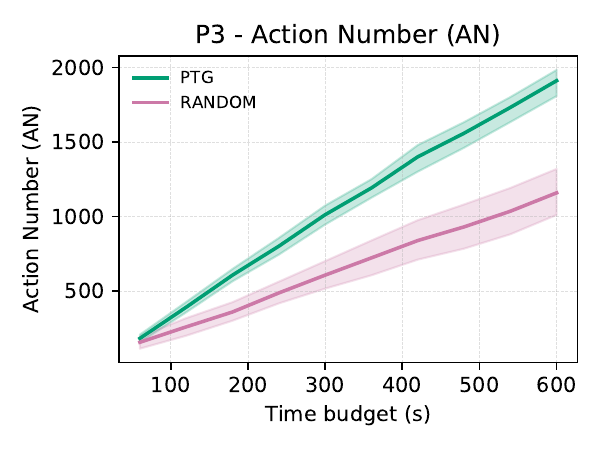}
        \captionsetup{skip=0pt}
        \caption{P3}
        \label{fig:rq2-p3-mean-an}
    \end{subfigure}
    \begin{subfigure}[t]{0.325\linewidth}
        \centering
        \includegraphics[width=\linewidth]{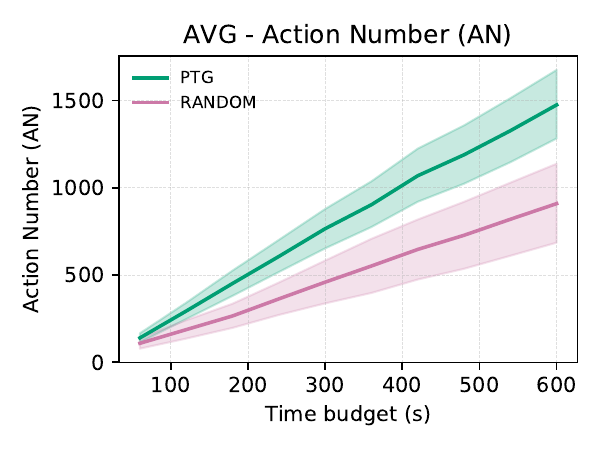}
        \captionsetup{skip=0pt}
        \caption{AVG}
        \label{fig:rq2-avg-mean-an}
    \end{subfigure}
    \captionsetup{skip=1pt}
    \caption{Action Number (AN) under different time budgets.}
    \label{fig:rq2-mean-an}
\end{figure}

\begin{figure}[t]
    \centering
    \begin{subfigure}[t]{0.325\linewidth}
        \centering
        \includegraphics[width=\linewidth]{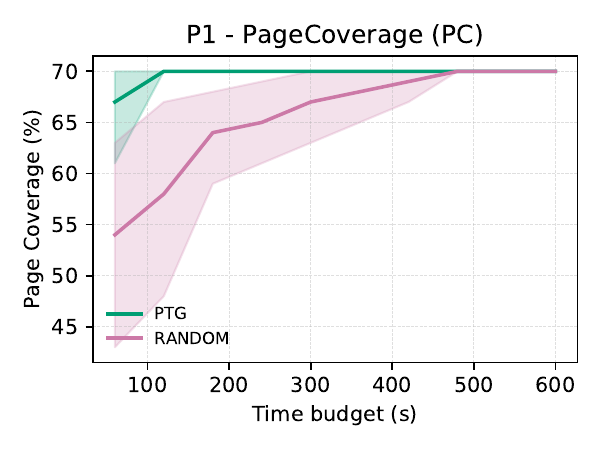}
        \captionsetup{skip=0pt}
        \caption{P1}
        \label{fig:rq2-p1-mean-pc}
    \end{subfigure}
    \begin{subfigure}[t]{0.325\linewidth}
        \centering
        \includegraphics[width=\linewidth]{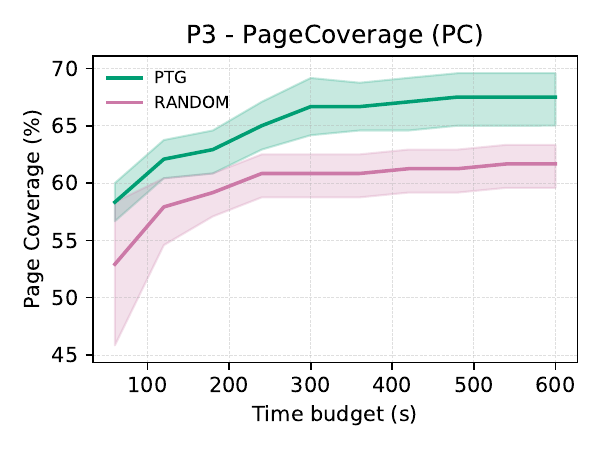}
        \captionsetup{skip=0pt}
        \caption{P3}
        \label{fig:rq2-p3-mean-pc}
    \end{subfigure}
    \begin{subfigure}[t]{0.325\linewidth}
        \centering
        \includegraphics[width=\linewidth]{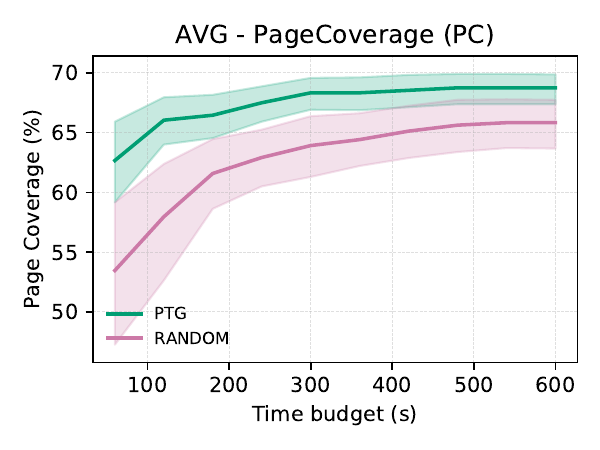}
        \captionsetup{skip=0pt}
        \caption{AVG}
        \label{fig:rq2-avg-mean-pc}
    \end{subfigure}
    \captionsetup{skip=1pt}
    \caption{Page Coverage (PC) under different time budgets.}
    \label{fig:rq2-mean-pc}
\end{figure}

\begin{figure}[t]
    \centering
    \begin{subfigure}[t]{0.325\linewidth}
        \centering
        \includegraphics[width=\linewidth]{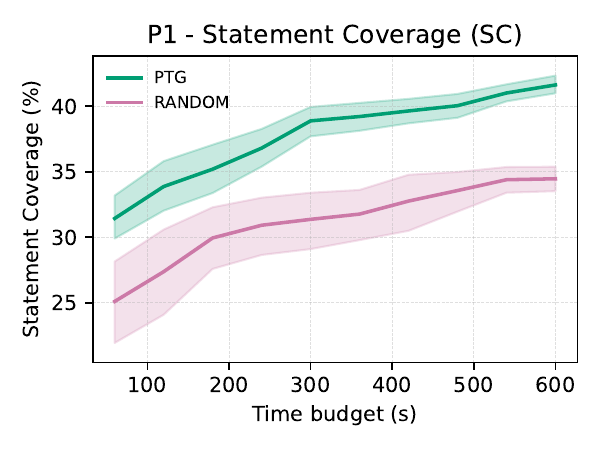}
        \captionsetup{skip=0pt}
        \caption{P1}
        \label{fig:rq2-p1-mean-sc}
    \end{subfigure}
    \begin{subfigure}[t]{0.325\linewidth}
        \centering
        \includegraphics[width=\linewidth]{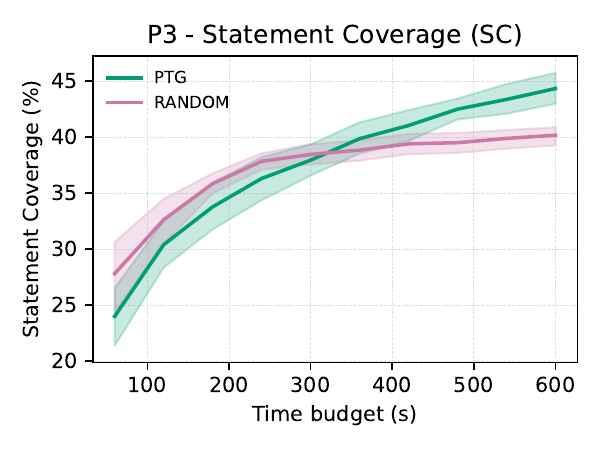}
        \captionsetup{skip=0pt}
        \caption{P3}
        \label{fig:rq2-p3-mean-sc}
    \end{subfigure}
    \begin{subfigure}[t]{0.325\linewidth}
        \centering
        \includegraphics[width=\linewidth]{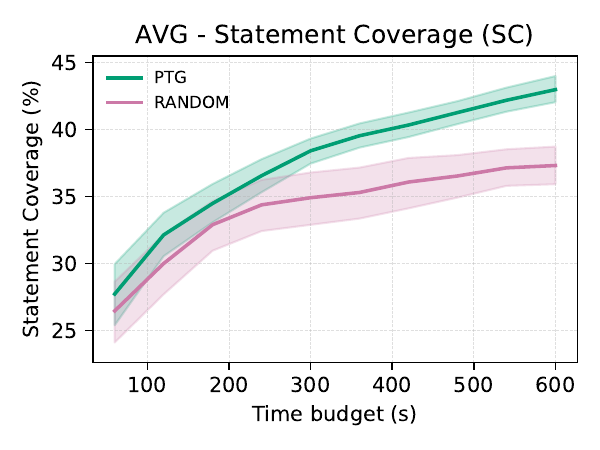}
        \captionsetup{skip=0pt}
        \caption{AVG}
        \label{fig:rq2-avg-mean-sc}
    \end{subfigure}
    \captionsetup{skip=1pt}
    \caption{Statement Coverage (SC) under different time budgets.}
    \label{fig:rq2-mean-sc}
\end{figure}

\begin{figure}
    \centering
    \begin{subfigure}[t]{0.325\linewidth}
        \centering
        \includegraphics[width=\linewidth]{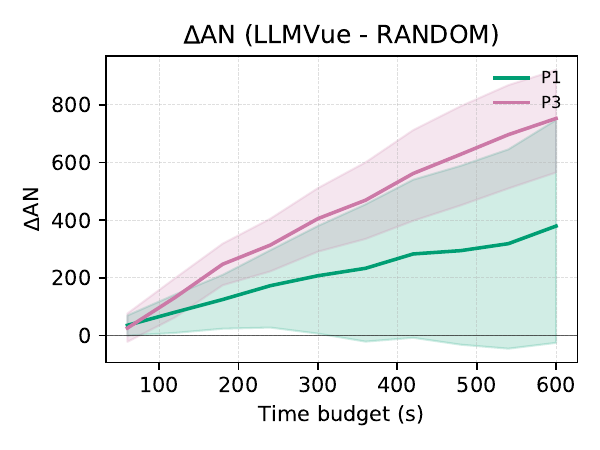}\captionsetup{skip=0pt}
        \caption{$\Delta$AN (LLMVue$-$Random)}
        \label{fig:rq2-delta-an}
    \end{subfigure}
    \begin{subfigure}[t]{0.3325\linewidth}
        \centering
        \includegraphics[width=\linewidth]{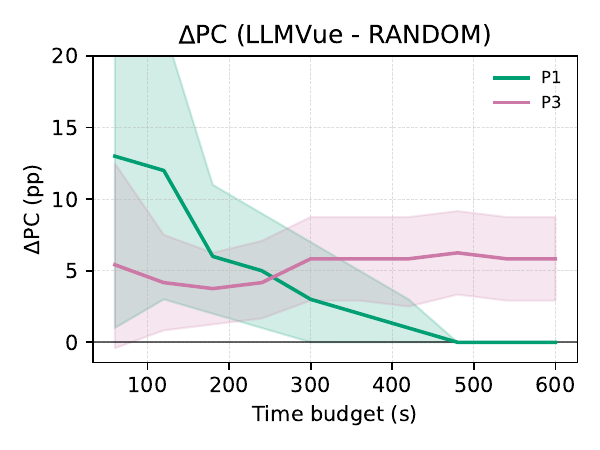}\captionsetup{skip=0pt}
        \caption{$\Delta$PC (pp)}
        \label{fig:rq2-delta-pc}
    \end{subfigure}
    \begin{subfigure}[t]{0.325\linewidth}
        \centering
        \includegraphics[width=\linewidth]{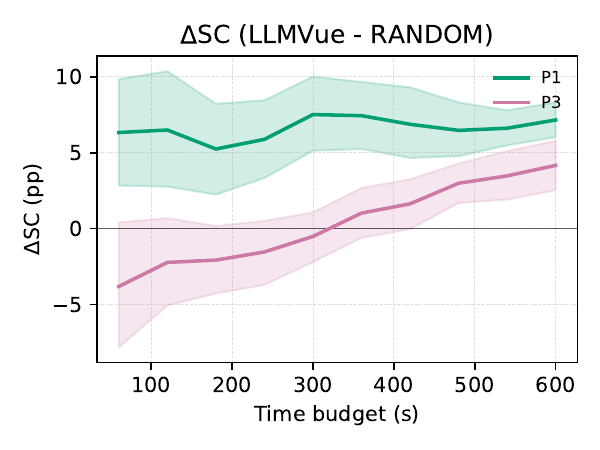}\captionsetup{skip=0pt}
        \caption{$\Delta$SC (pp)}
        \label{fig:rq2-delta-sc}
    \end{subfigure}\captionsetup{skip=1pt}
    \caption{Performance gaps between \textsc{LLMVue}-guided and random exploration.}
    \label{fig:rq2-delta-timeseries}
\end{figure}

\subsubsection{Observations}

Table~\ref{tab:rq2-timebudget} summarizes the testing effectiveness for three representative time budgets (60s/300s/600s).
It reports the mean values with 95\% confidence intervals (CIs) in the AN/PC/SC columns, and the corresponding improvements of \textsc{LLMVue}-guided exploration over the random baseline in the $\Delta$AN/$\Delta$PC/$\Delta$SC columns.
Figures~\ref{fig:rq2-mean-an}--\ref{fig:rq2-mean-sc} show the full time-series of AN/PC/SC for both projects (P1 and P3) and a pooled view across the two projects (AVG) view\footnote{Lines show mean values; shaded bands indicate 95\% CIs. AVG is a pooled view computed from runs in P1 and P3 under the same mode and time budget (20 runs per point). It is reported only as an aggregate summary, not a representativeness claim.}
Figure~\ref{fig:rq2-delta-timeseries} shows the performance improvements (LLMVue$-$Random) with 95\% CI bands\footnote{Lines show the mean difference (LLMVue) over 10 runs; shaded bands indicate 95\% bootstrap CIs of the mean differences.
Positive values mean that the \textsc{LLMVue}-guided exploration was more effective than the random baseline (under the same time budget).}.
For each project, mode, and time budget, the mean value, over 10 runs, is reported.
The 95\% CIs for the mean using a nonparametric percentile bootstrap: 
The 10 runs were resampled with replacement over 5000 iterations, with the 2.5th and 97.5th percentiles of the bootstrap mean distribution then taken.
The Figure~\ref{fig:rq2-delta-timeseries} bootstrap difference of means was calculated by independently resampling the PTG and Random runs at each budget, and computing $\Delta=\overline{x}_{PTG}-\overline{x}_{Rand}$ for each iteration.
The random seed was recorded for reproducibility. 
Based on the experimental data, the following observations can be made:

\begin{itemize}[leftmargin=2em, topsep=0pt, itemsep=0pt, parsep=0pt, partopsep=0pt]
    \item 
    \textbf{\textsc{LLMVue} substantially improves exploration efficiency (AN) under the same time budget.}
    As shown in Table~\ref{tab:rq2-timebudget} (AN and $\Delta$AN columns), \textsc{LLMVue}-guided testing consistently executes more actions than the random exploration baseline.
    For P1, the AN increases 
    from 60.1 to 95.8 at 60s ($+59.4\%$); 
    from 310.6 to 518.0 at 300s ($+66.8\%$); and
    from 658.8 to 1038.3 at 600s ($+57.6\%$).
    For P3, 
    The improvement is also evident for P3, including 
    from 606.5 to 1011.8 at 300s ($+66.8\%$), and from 1158.7 to 1911.1 at 600s ($+64.9\%$).
    This trend can also be seen in Figures~\ref{fig:rq2-p1-mean-an} and \ref{fig:rq2-p3-mean-an}, where the \textsc{LLMVue}-guided curve stays above the random baseline across budgets.
    Notably, the random baseline exhibits substantially larger run-to-run variance on AN (e.g., P1 at 300s: $310.6\pm186.7$ vs.\ PTG $518.0\pm32.2$), while PTG achieves both higher action counts and tighter CIs.
    This improvement translates into higher action throughput under the same budget (e.g., at 300s: 1.73 vs.\ 1.04 actions per second on P1, and 3.37 vs.\ 2.02 on P3).
    The pooled AVG view (Figure~\ref{fig:rq2-avg-mean-an}) indicates that this AN advantage exists across both projects.
    The \textsc{LLMVue}-guided exploration reduces run-to-run variance in AN (narrower CIs), indicating more stable exploration throughput.

    \item 
    \textbf{\textsc{LLMVue} improves page coverage early, and maintains advantages on the structurally complex project.}
    The PTG improves page coverage (PC) for low budgets.
    The P1 PC increases from 54\% to 67\% for 60s (+13pp; see the PC (\%) and $\Delta$PC columns in Table~\ref{tab:rq2-timebudget}), reaches 70\% by 120s, and remains stable through 600s.
    In contrast, the random baseline increases more gradually, and only reaches 70\% for larger budgets (480s--600s; Figure~\ref{fig:rq2-p1-mean-pc}).
    The P3 PTG yields sustained PC gains across budgets (e.g., 60.83\% to 66.67\% at 300s, and 61.67\% to 67.50\% at 600s).
    The Figure~\ref{fig:rq2-delta-pc} delta curves show that PC improvements are consistently positive over time, for both projects (although the gap naturally shrinks once the PC plateaus on P1).
    The pooled AVG PTG (Figure~\ref{fig:rq2-avg-mean-pc}) also achieves higher PC throughout the budget range, suggesting that structural guidance accelerates route-level discovery.
    The PC gap shrinks on P1 after 120s because PC saturates at 70\% for both modes, leaving limited potential for further gains.

    \item 
    \textbf{Statement coverage shows clear long-budget benefits on P1 and delayed benefits on P3 under \textsc{LLMVue}-guided exploration.} 
    The statement coverage (SC) improvements for P1 (highest F1) are substantial, and increase with the time budget (e.g., 31.37\% to 38.89\% at 300s; 34.47\% to 41.63\% at 600s).
    The delta SC curve becomes reliably positive from mid-level budgets onward (Figure~\ref{fig:rq2-delta-sc}), matching the widening gap in Figure~\ref{fig:rq2-p1-mean-sc}.
    The P3 SC (largest PTG) is slightly lower under PTG for lower budgets (e.g., 23.99\% vs.\ 27.80\% at 60s); 
    remains comparable around 300s (37.93\% vs.\ 38.44\%); and 
    becomes higher for larger budgets (44.32\% vs.\ 40.15\% at 600s).
    Consistent with this, the delta SC curve crosses from near-zero/negative to positive for larger budgets (e.g., positive from 480s onward in Figure~\ref{fig:rq2-delta-sc}).
    The pooled AVG view (Figure~\ref{fig:rq2-avg-mean-sc}) also shows that SC gains are less at low budgets, but that they become clearly positive as the budget grows:
    This indicates that navigation benefits increasingly translate into deeper code execution as exploration continues.
    A possible explanation for this is that \textsc{LLMVue}-guided exploration prioritizes structural navigation early (discovering more routes), which may delay deep-statement execution until larger budgets:
    In contrast, the random approach occasionally spends more time within a smaller set of routes and executes statements earlier.
\end{itemize}

\finding{2}{Under the same time budgets, \textsc{LLMVue}-guided exploration outperforms random exploration by executing more actions and reaching new routes earlier.
Coverage gains, especially for statement coverage, are more pronounced as the budget increases.}

\section{Discussion} 
\label{sec:discussion}

In this section, we discuss why \textsc{LLMVue} works, the implications for model-based GUI testing of Vue.js applications, when PTG guidance is less effective, and opportunities for improvement revealed by our experiments.

\subsection{Why \textsc{LLMVue} Works}

Our results suggest that a hybrid design that combines static analysis with LLM-based reasoning is effective for modeling Vue.js navigation.
Stage~I adopts a high-recall strategy, conservatively incorporating candidate navigation edges across router configurations, templates, and scripts.
Stage~II leverages an LLM to refine edge semantics (e.g., selector enhancement and condition extraction), which is difficult to achieve solely with rules in Vue/Element UI projects where navigation logic is frequently encoded in event handlers and UI library abstractions.
Stage~III further normalizes and deduplicates the constructed edges to reduce redundancy and improve model usability for downstream testing.
Consistent with our expectation, \textsc{LLMVue} achieves high precision and high F1-score across ten projects in RQ1, indicating that the three stages jointly balance completeness and correctness.

\subsection{Implications for Model-Based GUI Testing of Vue.js Applications}

Our study (RQ2) shows that the \textsc{LLMVue}-constructed PTG can serve as an effective structural model for guiding GUI exploration.
First, PTG guidance consistently improves exploration efficiency (AN), suggesting that the structural graph helps testers avoid ineffective interactions and repeatedly traversing the same UI states.
Second, PTG guidance improves page coverage (PC), especially at short budgets, which is valuable in time-constrained testing settings.
Third, the improvements in statement coverage (SC) depend on the time budget and the interaction structure.
For the project with the highest extraction quality (P1), PTG guidance provides a stable and consistent increase in statement coverage as the budget increases.
On the structurally complex project (P3), PTG guidance primarily accelerates route-level discovery, while deeper intra-page interactions are accumulated progressively; this explains the delayed SC improvement observed in longer budget phases.

\subsection{When PTG Guidance is Less Effective}

Our analysis also reveals scenarios where PTG guidance alone may not immediately improve code coverage.
For example, in the applications with intensive intra-page behaviors, a significant portion of statements is executed through widget-level interactions that do not trigger route changes.
In these cases, the main advantage of PTG guidance is faster access to relevant pages; however, systematically exploring the sequence of events within a page requires additional heuristics.
Furthermore, the effectiveness of PTG guidance depends on the completeness of the constructed PTG; conditional navigation and callback-driven navigation may be under-modeled, resulting in missing paths. These omissions prevent PTG guidance from prioritizing these paths, thus delaying access to some pages and reducing the effectiveness of the guidance.

\subsection{Opportunities for Improvement}

The above limitations suggest several potential future directions.
First, closing the loop between extraction and testing: If the predicted target page is not reached (or a page is consistently missed), the results can be fed back to $G_{\textsc{ext}}$ so that the graph can be optimized through targeted reanalysis of relevant components.
Second, role-aware and state-aware modeling: For applications with role-dependent menus and state-dependent routing, using a multi-context PTG (e.g., one subgraph per role) may be more effective than using a single global graph.
Third, selector disambiguation: Introducing additional DOM context (e.g., ancestor chains, ARIA tags, or stable component identifiers) can reduce selector ambiguity and further improve accuracy.
Fourth, hybrid exploration policies: Combining PTG-guided navigation with systematic in-page exploration (e.g., operation priority ranking based on coverage feedback) can better translate navigation advantages into improved code coverage.

Overall, our findings support the feasibility of using \textsc{LLMVue} to assist model-based GUI testing for Vue.js applications and identify key areas for future work to improve robustness and testing depth.

\subsection{Threats to Validity} 

In this section, we mainly discuss some potential threats to validity.

\subsubsection{Ground-Truth Construction}
Although $G_{\textsc{gt}}$ was built using a deterministic, version-locked expert protocol co-developed with Vue/Element UI practitioners, some residual ambiguity may still exist in specific corner cases, such as condition-controlled widgets and callback-driven routing.
We mitigate this threat through pre-annotation calibration, centralized arbitration for ambiguous cases, and automated sanity checks on selector validity and route consistency.

\subsubsection{Metrics for Model Quality and Testing Effectiveness}
The metrics in RQ1 (precision/recall/F1 and GED) and RQ2 (AN/PC/SC) capture edge-level structural fidelity and testing progress; however, they do not directly quantify robustness to UI variations or intra-page state diversity.
We mitigate this threat by applying a version-locked GT protocol with selector normalization and edge attributes (\texttt{selectorKind}, \texttt{condition}, and \texttt{note}), and reporting multiple complementary metrics rather than relying on a single indicator.

\subsubsection{Nondeterminism in Extraction and Execution}
Automated GUI exploration may exhibit run-to-run variance due to asynchronous rendering and network timing.
We mitigate this threat by using deterministic decoding for PTG extraction (GPT-4o with temperature = 0) and repeating each RQ2 configuration 10 times under a fixed environment.

\subsubsection{Generalizability of Projects and Settings}
The conclusions from our evaluation are limited to the context of Vue.js SPAs with Vue Router (as conducted on ten open-source projects), and may not generalize to other front-end frameworks or architectural patterns.
We partially mitigate this threat by selecting diverse projects, reporting detailed RQ2 results for two representative projects (P3 with the largest number of edges, and P1 with the highest F1), and by releasing a replication package to support follow-up evaluations.

\section{Conclusions and Future Work} 
\label{sec:conclusion} 

\subsection{Conclusions}
This paper investigated the feasibility of LLM-assisted model-based GUI testing for Vue.js web applications.
We proposed a new framework, namely \textsc{LLMVue}, which automatically extracts a global Page Transition Graph (PTG) from Vue.js source code and uses it as a structural model to guide automated GUI exploration.
\textsc{LLMVue} follows a three-stage pipeline: (i) high-recall static extraction from router configurations, templates, and scripts; (ii) LLM-based semantic refinement to enhance selectors and annotate conditional navigation; and (iii) graph normalization to consolidate redundant edges and improve model usability.

Our experiments on ten open-source Vue.js web applications demonstrate that \textsc{LLMVue} can effectively extract accurate PTGs, capturing the route-to-route navigation induced by user interactions across router configurations and component implementations.
Across all projects, the constructed PTGs achieve an average precision of 94.57\%, recall of 88.79\%, and F1-score of 91.25\%.
In addition, PTG-guided exploration improves testing effectiveness substantially, compared with the random exploration baseline under the same time budgets.
These improvements can be observed in both a high-quality constructed-model project (P1) and a structurally complex project (P3), helping mitigate potential project-selection bias. 
More specifically, on P1, \textsc{LLMVue} not only improves exploration efficiency, but also achieves consistently higher statement coverage at medium and long budgets.
With increased testing budgets on the structurally complex project P3, \textsc{LLMVue} achieves higher page coverage and, more importantly, achieves higher code coverage.

Overall, our findings suggest that \textsc{LLMVue}-constructed PTGs provide a practical structural abstraction for systematic GUI testing of Vue.js applications.

\subsection{Future Work}

Although the results demonstrate the feasibility and effectiveness of \textsc{LLMVue}, several opportunities remain for further improvement.

\subsubsection{Closing the Extraction–Testing Feedback Loop}
The current \textsc{LLMVue} workflow follows a source-driven process in which a PTG is first extracted from Vue.js source code and then used to guide GUI exploration.
However, the testing phase can provide useful runtime evidence about the quality of the constructed model.
For example, repeatedly failed actions, unreachable target pages, and consistently unexplored routes may indicate inaccurate selectors, missing transitions, or incomplete modeling of dynamic navigation behaviors.
Future work will investigate mechanisms for feeding execution traces and exploration outcomes back into PTG refinement.
Such a closed-loop process may help correct inaccurate edges, recover missing transitions, and improve the robustness of PTG-guided testing for evolving Vue.js applications.

\subsubsection{Role-Aware and State-Aware Modeling}
As discussed in this paper, navigation behavior in Vue.js applications is often influenced by user roles, authentication states, permissions, and other runtime conditions.
Although \textsc{LLMVue} currently preserves conditional information as annotations associated with PTG edges, the resulting graph remains a unified navigation model.
This abstraction may not fully capture context-dependent navigation structures, especially when different menu items or route transitions are exposed only under specific roles or states.
Future work will explore role-aware and state-aware extensions of PTGs that explicitly represent navigation behaviors under different execution contexts.
Such models may provide more precise guidance for GUI exploration and improve testing effectiveness for applications with complex access-control and state-dependent routing mechanisms.

\subsubsection{Coverage-Aware Intra-Page Exploration}
The experimental results show that PTG-guided exploration is effective at improving route discovery and page coverage.
However, a substantial portion of application functionality is exercised through widget-level interactions that do not trigger route transitions, such as form submissions, table operations, dialog interactions, and other intra-page behaviors.
This helps explain why improvements in page coverage do not always immediately translate into higher statement coverage, particularly under limited testing budgets.
Future work will investigate the integration of PTG-guided navigation with coverage-aware intra-page exploration strategies.
By combining structural route-level guidance with runtime coverage feedback and systematic widget exploration, future approaches may further improve statement coverage and better test widget-level behaviors in complex Vue.js web applications.



\section*{Acknowledgment}
This work is supported by the Science and Technology Development Fund of Macau, Macao SAR, under Grant Nos. 0069/2025/RIB2 and 0021/2023/RIA1.


{
\bibliographystyle{ACM-Reference-Format}
\bibliography{ref.bib}
}


\end{document}